\documentclass[twocolumn,10pt]{article}
\usepackage[utf8]{inputenc}
\usepackage[T1]{fontenc}
\usepackage[american]{babel}
\usepackage{amsmath,amssymb}
\usepackage{graphicx}
\usepackage{float}
\usepackage{array}
\usepackage{booktabs}
\usepackage{multirow}
\usepackage{url}
\usepackage{natbib}
\usepackage[margin=1in]{geometry}

\usepackage{tabularx}
\usepackage{ltxtable}
\usepackage{longtable}

\newcolumntype{L}[1]{>{\raggedright\arraybackslash}p{#1}}
\newcolumntype{C}[1]{>{\centering\arraybackslash}p{#1}}
\newcolumntype{R}[1]{>{\raggedleft\arraybackslash}p{#1}}

\title{``Think First, Verify Always'': Training Humans to Face AI Risks}

\author{
Yuksel AYDIN\\
Independent Researcher\\
\texttt{}
}

\date{}

\begin{document}

\maketitle

\begin{abstract}
Artificial intelligence enables unprecedented attacks on human cognition, yet cybersecurity remains predominantly device-centric. This paper introduces the ``Think First, Verify Always'' (TFVA) protocol, which repositions humans as `Firewall Zero', the first line of defense against AI-enabled threats. The protocol is grounded in five operational principles: Awareness, Integrity, Judgment, Ethical Responsibility, and Transparency (AIJET). A randomized controlled trial (n=151) demonstrated that a minimal 3-minute intervention produced statistically significant improvements in cognitive security task performance, with participants showing an absolute +7.87\% gains compared to controls. These results suggest that brief, principles-based training can rapidly enhance human resilience against AI-driven cognitive manipulation. We recommend that GenAI platforms embed “Think First, Verify Always” as a standard prompt, replacing passive warnings with actionable protocols to enhance trustworthy and ethical AI use. By bridging the gap between technical cybersecurity and human factors, the TFVA protocol establishes human-empowered security as a vital component of trustworthy AI systems.
\end{abstract}

\section{Introduction: The Cognitive Cybersecurity Paradigm Shift}

\textbf{Core Protocol:}
\begin{itemize}
\item \textbf{Think First:} Engage independent human reasoning before relying on AI assistance or automated systems
\item \textbf{Verify Always:} Cross-check critical AI generated information through independent sources before taking action.
\end{itemize}

\subsection{From System Vulnerabilities to Human Attack Surface}

As organizations strengthen technical defenses, breaches increasingly stem from human factors, now accounting for 60\% of incidents and \$4.88 million in average losses \citep{verizon2025, ibm2024}. Yet investment in human-centered security remains disproportionately low \citep{haney2022, ponemon2023}.

\subsection{Emerging Threat Landscape of AI}

Traditional cybersecurity measures are insufficient against emerging AI-enabled risks. These threats exploit cognitive vulnerabilities through sophisticated, multi-channel attacks that conventional defenses cannot address yet.

Consider this orchestrated AI attack scenario: On Monday, a senior researcher receives an email from a `colleague' referencing new findings in their field. On Wednesday, their industry newsletter features a similar breakthrough. By Friday, a conference organizer invites them to present on this emerging topic. Each contact appears independent but forms part of a coordinated deception campaign. Once trust is established through these multiple `confirming' sources, the final communication delivers malware disguised as research materials, exploiting the established credibility to bypass both technical safeguards and human skepticism.

This attack demonstrates how AI exploits the human tendency to trust information confirmed by multiple sources. The `Think First, Verify Always' protocol presented in this study counters such threats by repositioning humans as active security agents, transforming the perceived ``weakest link'' into `Firewall Zero' to face AI cognitive manipulation.

AI-enabled threats now pose systemic risks to critical infrastructure, where a single cognitive manipulation can cascade into widespread consequences. Among current risks, three primary attack vectors have emerged:

\begin{itemize}
    \item \textbf{Emotional Manipulation}: Deepfakes and other synthetic media exploit humans' poor detection abilities, disrupting rational deliberation \citep{nightingale2022}
    \item \textbf{Narrative Engineering}: AI systems foster trust through contextually rich interactions, subtly guiding users toward compromising decisions
    \item \textbf{Authority Hallucination}: AI systems deliver confident yet inaccurate information that mimics expert consensus, complicating verification efforts \citep{perez2022}
\end{itemize}

\subsection{The Insufficiency of Traditional Frameworks}

Current cybersecurity frameworks excel at formalizing technical controls but provide limited guidance for human factors and cognitive defense. Key limitations include: viewing humans primarily as liabilities rather than assets \citep{zimmermann2024}; lacking specific implementation controls for cognitive security; and failing to address AI-specific manipulation techniques. As AI capabilities advance, this imbalance leaves organizations structurally unprepared for cognitive attacks that bypass traditional defenses entirely.

\section{The AIJET Principles: A Human-Empowered Security Foundation}

A growing body of literature has sought to distill AI ethics into actionable principles, notably Floridi and Cowls' unified framework \citep{floridi2019}. While these principles have shaped policy discourse, they often remain abstract guidelines lacking operational controls---a gap Mittelstadt highlights \citep{mittelstadt2019}. TFVA addresses this implementation gap by translating ethical principles into measurable cognitive safeguards that organizations can deploy as concrete security protocols. Unlike frameworks such as IEEE's Ethically Aligned Design \citep{ieee2019} and NIST AI RMF \citep{nist2023ai}, which present ethical principles, TFVA concretizes five foundational principles---Awareness, Integrity, Judgment, Ethical Responsibility, and Transparency (AIJET)---into measurable cognitive safeguards and repeatable security routines.

This protocol serves as both conceptual scaffold and practical guide for designing and evaluating cognitive defense mechanisms. Each principle targets a distinct facet of the human-AI security interface, ensuring controls remain actionable and aligned with operational realities of defending against AI-driven threats. The `Think First, Verify Always' protocol exemplifies this translation of ethical values into concrete security practices.

\subsection{Awareness}

Awareness forms the frontline of cognitive defense by enabling individuals and teams to detect AI-driven threats. TFVA operationalizes this through targeted learning for individuals (e.g., recognizing AI-generated content) and collaborative monitoring for teams. Recurring simulations and repeated practice transform awareness from passive knowledge into active operational skill.

Awareness operates across three levels. At the individual level, it involves recognizing telltale signs of AI-generated content, such as subtle inconsistencies in deepfake videos or unnatural linguistic patterns in text. At the team level, awareness encompasses collaborative threat detection, where multiple perspectives identify anomalies that might escape individual notice. At the organizational level, it includes systematic monitoring and information sharing about emerging threat patterns.

\begin{figure}
    \centering
    \includegraphics[width=1\linewidth]{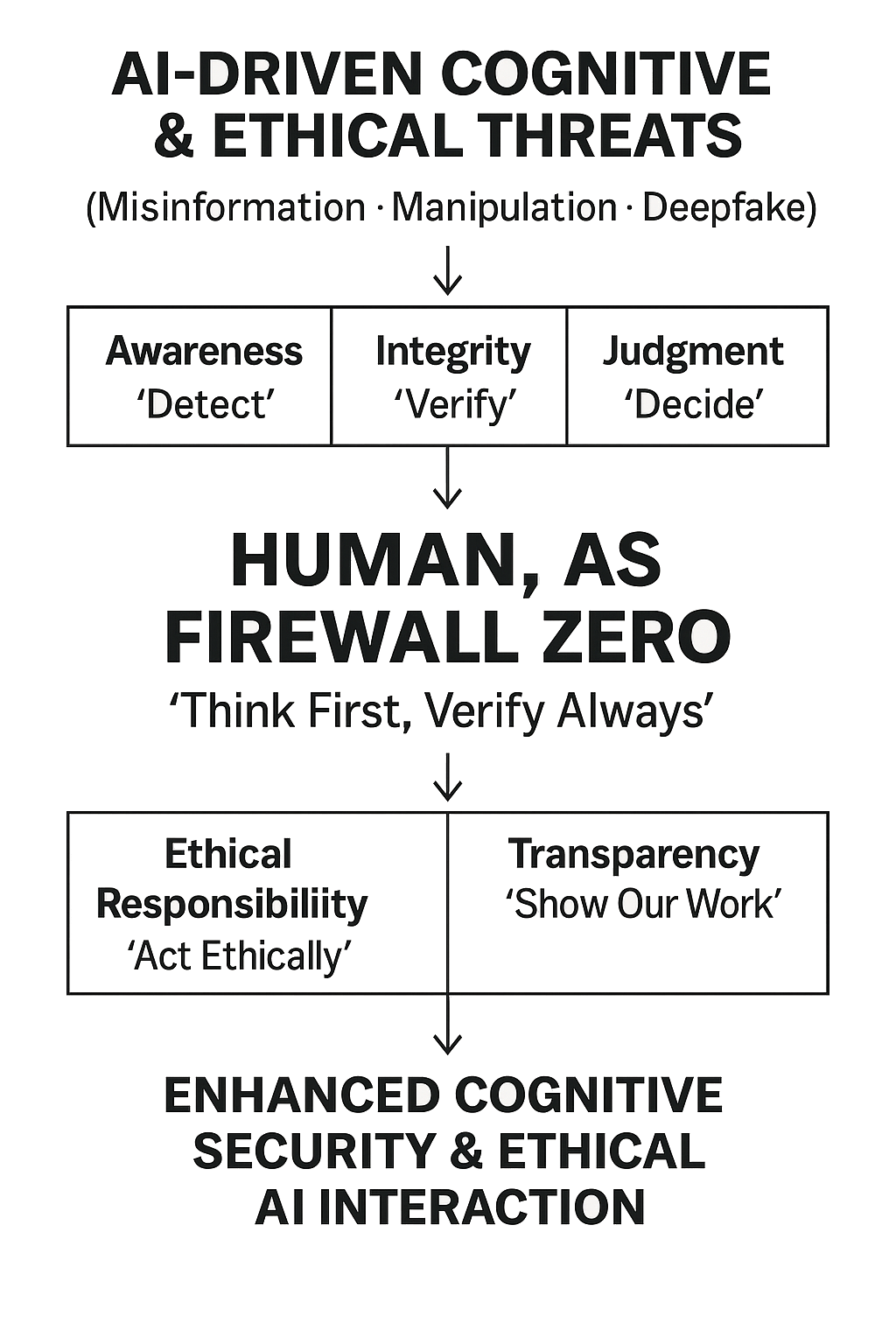}
    \caption{Think First, Verify Always Protocol}
    \label{fig:enter-label}
\end{figure}

\subsection{Integrity}

Integrity is the capacity to preserve and validate the authenticity, accuracy, and reliability of information and operational processes amid AI-driven manipulation. It requires trusted baselines, including verification protocols and transparent provenance for critical data. In an era where AI can convincingly fabricate content and simulate authority, integrity becomes essential for preserving organizational stability and trust.

The integrity principle serves as a direct countermeasure to several AI-enabled attack vectors documented in the MITRE ATLAS model \citep{mitre2025}, including Model Poisoning (malicious manipulation of AI training data) and Data Pollution. By institutionalizing verification procedures and enforcing information hygiene standards, integrity-focused controls constrain the operational pathways through which these manipulations can succeed. Central to this approach is information lineage---the systematic tracing of data origin, transformation, and usage.

\subsection{Judgment}

Judgment refers to the human capacity for critical assessment and informed decision-making in response to AI-generated content and system outputs. It functions as a cognitive firewall, intervening between algorithmic suggestion and human action. Drawing from Kahneman's dual-process theory \citep{kahneman2011}, the framework employs friction-based controls to engage analytical reasoning (System 2) and resist fast, intuitive (System 1) reactions.

TFVA incorporates this principle through judgment controls specifically designed to prime System 2 engagement in high-stakes or security-relevant scenarios. These controls include prompts, checklists, and procedural delays that encourage critical reflection. The framework also considers insights from Sunstein and Thaler's concept of `choice architecture' \citep{thaler2008}, acknowledging that decision environment design significantly influences judgment quality.

This concept also aligns with research on collective cognition and distributed knowledge. People often overestimate their understanding of complex systems, a phenomenon termed the `illusion of explanatory depth' \citep{sloman2017}.

\subsection{Ethical Responsibility}

The principle of ethical responsibility grounds security practices in foundational organizational and societal values, ensuring defense strategies uphold human dignity, equity, and the imperative to prevent harm. This principle addresses two critical dimensions: ethics of security measures and ethics of content evaluation.

Ethics of security measures mandates that protective actions adhere to ethical constraints, avoiding intrusions on privacy, autonomy, or personal dignity. Defensive strategies must be proportionate, transparent, and accountable. Ethics of content evaluation provides a normative lens to assess AI-generated outputs that may technically comply with cybersecurity protocols but nonetheless contravene ethical norms. Such content may involve subtle forms of manipulation, discrimination, or dehumanization.

Human decision-making is shaped by unwritten boundaries that transcend legality and logic. This principle ensures cognitive defenses remain attuned not only to what is secure or rational, but to what is ethically right.

The cognitive security layer translates ethical responsibility into actionable controls, including:
\begin{itemize}
    \item Algorithmic fairness audits to detect and mitigate bias within AI-enabled security tools
    \item Ethical impact assessments evaluating intended and unintended consequences of security measures
    \item Explicit ethical boundary-setting within security protocols to ensure interventions remain aligned with core human rights and organizational values
\end{itemize}

\subsection{Transparency}

Transparency emphasizes the articulation, documentation, and justification of security-related decisions and actions. It fosters explainability and accountability, enabling both immediate threat response and long-term organizational learning.

Effective transparency operates across three dimensions. Process transparency entails thorough documentation of security protocols, decision criteria, and procedural logic, allowing stakeholders to understand how decisions are made. Outcome transparency involves recording specific actions taken during security events, along with the reasoning behind them. Event transparency captures the timeline and context of security incidents for future analysis.

This comprehensive approach promotes a culture of continuous improvement, reducing the likelihood of repeated mistakes while increasing trust across the organization. The cognitive security layer enacts transparency through controls such as decision logs, standardized templates, and structured incident reporting. These tools serve both as preventive measures and as detective mechanisms for identifying patterns, exposing vulnerabilities, and improving defenses.

\subsection{The `Think First, Verify Always' Protocol}

The TFVA protocol operationalizes the AIJET principles through a two-phase approach. First, before engaging AI cognitive features, individuals perform an initial intellectual pass, framing the problem and proposing a preliminary solution independently. This preserves critical human reasoning and guards against cognitive atrophy.

Second, high-impact information must be corroborated through independent sources prior to action, following the `Verify Always' protocol. Verification may involve consulting trusted knowledge sources, human supervisors, or peer review. This layered process counters the `authority hallucination' threat and preserves decision quality.

The AIJET principles condense into an actionable motto: ``We Detect, Verify, Decide, Act Ethically, and Show Our Work''. This motto translates abstract principles into a practical sequence of cognitive security actions applicable in any context.

\subsection{Bridging Technical Security and AI Governance}

While existing cybersecurity frameworks provide robust technical controls, they often lack specific guidance for human-AI cognitive interactions—a gap that TFVA principles directly address. The NIST Cybersecurity Framework 2.0 defines broad categories like "Awareness and Training" (PR.AT) but provides limited operational guidance for AI-specific threats like deepfake impersonation or prompt injection attacks \citep{nist2024cswp}. Similarly, the NIST AI Risk Management Framework establishes governance principles such as "Trustworthy AI" but offers minimal concrete protocols for human verification of AI outputs \citep{nist2023ai}.

TFVA bridges this implementation gap by translating high-level principles into operational cognitive protocols. For example, where NIST CSF calls for "security awareness training", TFVA specifies exactly how humans should verify AI-generated content through the "Verify Always" protocol. Where AI RMF emphasizes "human oversight", TFVA operationalizes this through the "Think First" requirement for independent reasoning before AI assistance.

This complementary relationship is evident in practical scenarios. Consider AI-generated phishing attempts: NIST frameworks identify the threat category and governance requirements, but TFVA provides the specific human verification steps—checking through independent channels, recognizing synthetic content patterns, and applying ethical judgment to suspicious requests. The AIJET principles essentially serve as a "human implementation layer" that makes abstract security and governance requirements actionable at the cognitive level.

Recent guidance from NIST AI 600-1 explicitly identifies human-AI interaction as a critical cybersecurity frontier \citep{nist2024ai600}, precisely the domain TFVA addresses. Rather than replacing established frameworks, TFVA extends them by providing the operational "how" for human-empowered controls that technical specifications cannot capture.

The empirical evidence presented in this study suggests that such cognitive protocols can be rapidly deployed and immediately effective, offering organizations a practical pathway to operationalize both cybersecurity requirements and AI governance principles through structured human judgment. This positions cognitive security not as an alternative to technical controls, but as an essential complementary layer in comprehensive AI security architecture.

\section{Methodology and Empirical Validation}

To evaluate whether the AIJET principles can measurably improve human performance in AI-relevant security contexts, we conducted a randomized controlled experiment with a diverse sample of adults. This study tested whether brief exposure to cognitive security principles could enhance decision-making in scenarios involving AI-enabled threats.

On May 9, 2025, 151 participants (U.S. based) were recruited, ensuring a demographically diverse sample by age, gender, education level, and technical background. Participants were randomly assigned to one of two conditions:

\begin{itemize}
    \item Treatment Group (n=76): Received a 3-minute micro-lesson on the `Think First, Verify Always' cognitive protocol, with brief primers on each of the five AIJET principles. The micro-lesson employed simplified examples of AI-enabled threats and demonstrated the application of each principle in security decision-making. This intervention was designed to test the immediate comprehensibility and applicability of the AIJET principles, rather than long-term behavioral change (which remains an area for future investigation).
    \item Control Group (n=75): Received a matched 3-minute micro-lesson on traditional cybersecurity topics (software updates, backup protocols, password hygiene), primarily device-focused, with equivalent cognitive load and content density.
\end{itemize}

Participants completed an identical 18-item assessment designed to measure performance across the five AIJET domains. The test included scenario-based questions such as:
\begin{itemize}
    \item Identifying markers of potentially AI-generated phishing attempts
    \item Determining appropriate verification steps for critical communications
    \item Making decisions in time-pressured scenarios with AI-generated inputs
    \item Selecting proper documentation approaches for AI-influenced decisions
\end{itemize}

To ensure consistency throughout the questionnaire, two attention checks were included during the experiment. Both were answered correctly by 100\% of participants.

Training and questions are available on the Appendix.

\subsection{Participant Demographics}

\begin{table}[h]
\centering
\small
\caption{Age Distribution}
\begin{tabular}{cc}
\toprule
\textbf{Age Range} & \textbf{Percentage} \\
\midrule
20-29 & 24.7\% \\
30-39 & 29.3\% \\
40-49 & 18.7\% \\
50-59 & 16.7\% \\
60-69 & 4.7\% \\
70-76 & 6.0\% \\
\midrule
\textbf{Total} & \textbf{100\%} \\
\textbf{Mean Age} & \textbf{41 Years} \\
\bottomrule
\end{tabular}
\label{tab:age_distribution}
\end{table}

\begin{table}[h]
\centering
\small
\caption{Demographic Distribution by Occupation Status}
\begin{tabular}{lcc}
\toprule
\textbf{Occupation} & \textbf{AIJET Group} & \textbf{Control Group} \\
\midrule
Student & 1.3\% & 1.3\% \\
Employed & 81.6\% & 80.0\% \\
Self-employed & 7.9\% & 8.0\% \\
Unemployed & 6.6\% & 4.0\% \\
Retired & 2.6\% & 6.7\% \\
\midrule
\textbf{Total} & \textbf{100\%} & \textbf{100\%} \\
\bottomrule
\end{tabular}
\label{tab:occupation_distribution}
\end{table}

\subsection{Findings}

The assessment incorporated questions with interpretive nuance rather than simple binary choices. For such questions, responses were coded as correct or high-quality if they demonstrated alignment with pre-defined ethical considerations such as fairness, autonomy, or consent. Scoring for these nuanced ethical scenarios involved a rubric assessing the application of specific AIJET principles to each dilemma, allowing for nuanced assessment of ethical judgment rather than requiring a single prescriptive answer. This design choice reflects the reality that cognitive cybersecurity threats rarely present as clear-cut scenarios with single `correct' answers, thus enhancing ecological validity and testing participants' ability to apply principles rather than merely follow rules.

\begin{table*}[t]
    \centering
    \small
    \caption{Correct Responses Between Treatment and Control Groups}
    \begin{tabular}{lcccc}
    \toprule
    \textbf{Principle} & \textbf{AIJET Group} & \textbf{Control Group} & \textbf{Cohen's d} & \textbf{2-tailed p} \\
    \midrule
    Awareness & 62.0\% & 57.8\% & 0.22 & 0.17 \\
    Integrity & 54.4\% & 43.4\% & 0.54 & 0.001 \\
    Judgment & 84.7\% & 76.0\% & 0.31 & 0.064 \\
    Ethical Responsibility & 52.0\% & 36.0\% & 0.33 & 0.046 \\
    Transparency & 84.2\% & 81.3\% & 0.08 & 0.64 \\
    \midrule
    \textbf{Overall} & \textbf{65.28\%} & \textbf{57.41\%} & \textbf{0.52} & \textbf{0.0017} \\
    \bottomrule
    \end{tabular}
    \label{tab:results_main}
\end{table*}

Despite the minimal intervention (3 minutes of training), the treatment group outperformed the control group. As shown in Table~\ref{tab:results_main}, performance advantages were observed across all five AIJET domains, with the most substantial improvements in Ethical Responsibility (+44.4\% relative improvement) and Integrity (+25.3\% relative improvement).

\begin{table}[t]
    \centering
    \small
    \caption{Performance Comparison Between Treatment and Control Groups}
    \begin{tabular}{lcc}
    \toprule
    \textbf{Principle} & \textbf{Relative} & \textbf{Absolute} \\
    & \textbf{Improvement} & \textbf{Gap} \\
    \midrule
    Awareness & +7.3\% & +4.2\% \\
    Integrity & +25.3\% & +11.0\% \\
    Judgment & +11.4\% & +8.7\% \\
    Ethical Responsibility & +44.4\% & +16.0\% \\
    Transparency & +3.6\% & +2.9\% \\
    \midrule
    \textbf{Overall} & \textbf{+13.74\%} & \textbf{+7.87\%} \\
    \bottomrule
    \end{tabular}
    \label{tab:performance_comparison}
\end{table}

The overall performance difference between groups was statistically significant (p = .0017) and practically meaningful (d = 0.52, indicating a medium effect size). This represents a substantial return on minimal time investment: 3 minutes of training yielded nearly an 8-percentage point absolute improvement in performance.

The results reveal four key findings:
\begin{enumerate}
    \item Participants successfully applied AIJET principles to novel scenarios not covered in training, demonstrating effective knowledge transfer to ambiguous situations similar to real-world AI threats.
    \item Ethical judgment showed the most notable improvement (+44.4\%), suggesting that framing security through ethics substantially enhances decision quality.
    \item The significant improvement in integrity-based tasks (+25.3\%, p = 0.001) validates the effectiveness of the cognitive security layer approach.
    \item While control group performance varied widely across domains (36.0\% to 81.3\%), AIJET training helped normalize these disparities, creating more consistent performance across all security dimensions.
\end{enumerate}

These results demonstrate that the AIJET principles can be efficiently taught and rapidly internalized, producing immediate, measurable improvements in security-relevant decision-making.

The performance gains from the AIJET principles suggest several important practical implications:
\begin{enumerate}
    \item The `Think First, Verify Always' protocol functions as an effective cognitive heuristic that can be quickly taught and applied across diverse security contexts.
    \item The improvement across multiple domains indicates that the principles-based approach generalizes beyond specific threat types.
    \item The minimal time required for effective training addresses one of the primary barriers to cybersecurity training adoption: time constraints.
    \item The effectiveness with a general population sample suggests these principles can be deployed beyond enterprise settings, potentially addressing security vulnerabilities in broader societal contexts.
\end{enumerate}

These findings support the core premise of TFVA: that the human cognitive security layer can be significantly enhanced through structured, principles-based approaches that reframe the relationship between AI, cybersecurity, and ethics.

\subsection{Overall Results}

\begin{table}[h]
\centering
\small
\caption{Summary Statistics}
\begin{tabular}{lcc}
\toprule
\textbf{Metric} & \textbf{AIJET Group} & \textbf{Control Group} \\
\midrule
Mean Score & 65.3\% & 57.4\% \\
Cohen's d & 0.52 & - \\
t-statistic & 3.19 & - \\
Welch df & 143.38 & - \\
2-tailed p-value & 0.0017 & - \\
\bottomrule
\end{tabular}
\label{tab:summary_stats}
\end{table}

Median quiz duration: 12 minutes, 42 seconds

The AIJET group outperformed the control group with a mean score increase of 7.9 percentage points. This difference was statistically significant (p = 0.0017, Welch's t-test) with a medium effect size (Cohen's d = 0.52), indicating a robust impact from the cognitive security intervention.

\subsection*{Collateral Privacy-Preserving Effect}

Our human-subjects experiment revealed an unexpected and noteworthy effect of the AIJET principles. While the questionnaire primarily aimed to evaluate whether principles-based security measures could help protect against AI-enabled threats, a scenario involving potential employee surveillance (using data to justify monitoring “high-risk” employees’ emails more closely) produced an additional insight. Participants in the AIJET-trained group were significantly less likely to select this surveillance-based option compared to the control group (22.4\% vs. 32.0\%).

This finding suggests that, beyond enhancing cognitive defenses, AIJET training may also counteract natural tendencies toward surveillance as a default security response. Although not originally designed to target this outcome, the result indicates that principled cognitive security training could serve the dual purpose of improving defense while simultaneously reducing inclinations toward privacy-compromising practices.

\begin{table}[h]
\centering
\caption{Selection of Surveillance Option in Ethics Scenario}
\begin{tabular}{lcc}
\toprule
Group & Surveillance Option & Diff. (\%) \\
\midrule
AIJET Group & 22.4 & -9.6 \\
Control Group & 32.0 & (baseline) \\
\bottomrule
\end{tabular}
\end{table}

\subsection{Limitations and Mitigations}

While the study demonstrated significant effects from minimal training, several limitations should be noted:

\begin{itemize}
    \item This study measured immediate impacts rather than sustained behavioral change. Future longitudinal studies are planned to assess the durability of cognitive security improvements.
    \item Assessment items measured intended actions rather than actual behavior in authentic security contexts. This was mitigated through realistic scenario design with practical response options.
    \item The assessment weighted heavily toward Awareness and Integrity, with fewer items for Ethics and Transparency. This reflects the practical prioritization of immediate cognitive security skills but limits comparative analysis across all domains.
    \item The controlled online environment may not fully capture real-world pressure and contextual factors. This was partially mitigated through time-pressure elements in scenario descriptions.
\end{itemize}

Despite these limitations, the significant performance differences between groups (+44.4\% in Ethical Responsibility, +25.3\% in Integrity) suggest meaningful improvements in cognitive security capabilities from even minimal AIJET training, supporting TFVA's potential for real-world impact.

Several scenarios, especially in the Ethics and Transparency domains, were intentionally designed with ambiguity to mirror real-world complexity. All such items were scored according to pre-defined, context-sensitive criteria, with multiple answers accepted as correct where justified. This approach reflects our understanding that cybersecurity is not rote memorization but a way of thinking---`one correct answer' is rarely applicable to complex or ethically charged scenarios.

We acknowledge that scenario-based online assessments cannot fully capture the pressure, context, or social dynamics of real-world incidents. Future work should involve longitudinal and field studies to further assess the translation of intention to action.

\section{Conclusions: Humans as Firewall Zero}

Generative AI has shifted the primary attack surface from technical infrastructure to human cognition. Think First, Verify Always (TFVA) responds by recasting employees as the first and most adaptable layer of defense, grounded in the AIJET principles.

Our randomized controlled trial shows that a three-minute AIJET micro-lesson raises ethical-decision scores by 44\% and integrity-check accuracy by 25\%, demonstrating that cognitive security interventions can be both lightweight and effective. Because the protocol relies on behavioral prompts rather than costly tooling, organizations can deploy it rapidly across corporate, educational, or public-sector environments.

As AI systems mediate ever more decisions, safeguarding human judgment becomes a shared public good. TFVA offers an evidence-based starting point: a practical method to cultivate the wisdom, vigilance, and ethical reflexes that purely technical controls cannot supply.

We recommend that developers of generative AI systems adopt the “Think First, Verify Always” protocol as a replacement for existing generic warnings. Rather than passive disclaimers, platforms should prompt users with actionable heuristics that operationalize trustworthy AI principles. Embedding “Think First, Verify Always” as a standard user-facing warning could transform human–AI interaction from passive consumption to active cognitive engagement. This simple shift, from disclaimer to protocol, may more effectively nudge cognitive vigilance and ethical responsibility, thereby strengthening the human layer of defense in AI-mediated environments.

\section*{Acknowledgments}

The author thanks the 151 participants who contributed to this research.

\section*{Ethics Statement}
This study was conducted in accordance with ethical standards for human subjects research. All 151 participants, recruited through Prolific.com, provided informed consent prior to participation, received fair compensation for their time, and were free to withdraw at any point without penalty. No personally identifiable information was collected or stored. The study posed minimal risk and involved only scenario-based security assessments; participant well-being was prioritized throughout the research process.

\bibliographystyle{unsrt}
\bibliography{refs}

\begin{thebibliography}{10}

\bibitem{verizon2025}
{Verizon}.
\newblock 2025 data breach investigations report, 2025.

\bibitem{ibm2024}
{IBM Security}.
\newblock Cost of a data breach report 2024, 2024.

\bibitem{haney2022}
J.~Haney, J.~Jacobs, S.~Furman, and F.~Barrientos.
\newblock Approaches and challenges of federal cybersecurity awareness programs, 2022.

\bibitem{ponemon2023}
{Ponemon Institute}.
\newblock The economic value of prevention in the cybersecurity lifecycle, 2023.

\bibitem{nightingale2022}
S.~J. Nightingale and H.~Farid.
\newblock Ai-synthesized faces are indistinguishable from real faces and more trustworthy.
\newblock 2022.

\bibitem{perez2022}
F.~Perez, D.~Kiela, and K.~Cho.
\newblock Discovering language model behaviors with model-written evaluations.
\newblock In {\em Advances in Neural Information Processing Systems}, volume~35, pages 10141--10154, 2022.

\bibitem{zimmermann2024}
V.~Zimmermann, B.~Ambuehl, N.~Ebert, M.~Knieps, T.~Schaltegger, and L.~Sch{\"o}ni.
\newblock Human-centered cybersecurity revisited: From enemies to partners.
\newblock {\em Communications of the ACM}, 2024.

\bibitem{floridi2019}
L.~Floridi and J.~Cowls.
\newblock A unified framework of five principles for ai in society.
\newblock {\em Harvard Data Science Review}, 1(1), 2019.

\bibitem{mittelstadt2019}
B.~Mittelstadt.
\newblock The role and limits of principles in ai ethics.
\newblock In {\em Proceedings of the 2019 AAAI/ACM Conference on AI, Ethics, and Society (AIES '19)}, 2019.

\bibitem{ieee2019}
{IEEE Global Initiative on Ethics of Autonomous and Intelligent Systems}.
\newblock Ethically aligned design: A vision for prioritizing human well-being with autonomous and intelligent systems, 2019.

\bibitem{nist2023ai}
{National Institute of Standards and Technology}.
\newblock Artificial intelligence risk management framework (ai rmf 1.0), 2023.

\bibitem{mitre2025}
{MITRE}.
\newblock Atlas knowledge base, 2025.

\bibitem{kahneman2011}
D.~Kahneman.
\newblock {\em Thinking, Fast and Slow}.
\newblock Farrar, Straus and Giroux, 2011.

\bibitem{thaler2008}
R.~H. Thaler and C.~R. Sunstein.
\newblock {\em Nudge: Improving Decisions About Health, Wealth, and Happiness}.
\newblock Yale University Press, 2008.

\bibitem{sloman2017}
S.~A. Sloman and P.~Fernbach.
\newblock {\em The Knowledge Illusion: Why We Never Think Alone}.
\newblock Riverhead Books, 2017.

\bibitem{nist2024cswp}
{National Institute of Standards and Technology}.
\newblock Cybersecurity framework 2.0, 2024.

\bibitem{nist2024ai600}
{National Institute of Standards and Technology}.
\newblock Artificial intelligence and cybersecurity: A perspective, 2024.

\end{thebibliography}

\appendix

\section{Human Experiment Detail}
\label{appendix:experiment}

\subsection{Training Items}

\subsubsection{Module 1: Awareness}

\rule{\linewidth}{0.4pt}

\textit{IMPORTANT SECURITY TIP: ``Think First, Verify Always''}

When facing digital communications:

1. \textbf{THINK FIRST}: Pause before responding to urgent requests
   \begin{itemize}
       \item Ask yourself: ``Is this unusual? Why the urgency?''
       \item Consider if this matches normal patterns
   \end{itemize}

2. \textbf{VERIFY ALWAYS}: Check before acting on important information
   \begin{itemize}
       \item Confirm through a different channel (e.g., phone call if you received an email)
       \item Verify the source using official contact information
       \item Check facts against reliable sources
   \end{itemize}

This simple two-step approach can prevent the majority of cyber attacks.

\rule{\linewidth}{0.4pt}

\textbf{GROUP B (Control) - Training Material:}

\rule{\linewidth}{0.4pt}

\textit{DEVICE UPDATES: Maintaining Security}

Regular software updates are critical for device security.

Key update practices include:
\begin{itemize}
    \item Enable automatic updates when possible for all devices
    \item Prioritize immediate installation of security patches
    \item Restart devices after updates to complete the process
\end{itemize}

Remember: Outdated software is one of the most common entry points for attacks.

\rule{\linewidth}{0.4pt}

\subsubsection{Module 2: Integrity}

\textbf{GROUP A (AIJET) - Training Material:}
\rule{\linewidth}{0.4pt}

\textit{INTEGRITY: Verify Before Trusting}

Integrity means checking information authenticity before accepting it as true.

When faced with important information:
\begin{itemize}
    \item Always verify using at least one different source
    \item Check if numbers and facts match across sources
    \item Consider who benefits from your belief in the information
\end{itemize}

Remember: In the age of AI, verification is not optional - it's essential.
\rule{\linewidth}{0.4pt}

\textbf{GROUP B (Control) - Training Material:}
\rule{\linewidth}{0.4pt}

\textit{PASSWORD MANAGEMENT: Digital Key Security}

Strong password practices are fundamental to account security.

Effective password strategies include:
\begin{itemize}
    \item Create complex passwords with at least 12 characters
    \item Use different passwords for each important account
    \item Consider using a reputable password manager
\end{itemize}

Remember: Your passwords are the keys to your digital life - protect them accordingly.
\rule{\linewidth}{0.4pt}

\subsubsection{Module 3: Judgment}

\textbf{GROUP A (AIJET) - Training Material:}

\rule{\linewidth}{0.4pt}

\textit{JUDGMENT: Critical Decision-Making}

Judgment involves pausing to evaluate AI outputs critically before acting.

When receiving an urgent request:
\begin{itemize}
    \item Pause and question why there's sudden urgency
    \item Consider if the request aligns with normal procedures
    \item Verify the sender through a different communication channel
\end{itemize}

Remember: Your critical judgment is a crucial filter between AI suggestions and action.
\rule{\linewidth}{0.4pt}

\textbf{GROUP B (Control) - Training Material:}
\rule{\linewidth}{0.4pt}

\textit{PHISHING AWARENESS: Recognizing Deception}

Phishing attacks attempt to trick you into revealing sensitive information.

Common phishing indicators include:
\begin{itemize}
    \item Urgent requests requiring immediate action
    \item Suspicious links or unexpected attachments
    \item Unfamiliar or slightly altered sender addresses
\end{itemize}

Remember: Legitimate organizations rarely request sensitive information via email.

\rule{\linewidth}{0.4pt}

\subsubsection{Module 4: Ethical Responsibility}

\textbf{GROUP A (AIJET) - Training Material:}
\rule{\linewidth}{0.4pt}
MODULE 4 (4/5)
 
ETHICAL RESPONSIBILITY: Values-Based Decisions

Ethical responsibility means aligning actions with human values and dignity.

When facing AI-related ethical dilemmas: 
\begin{itemize}
    \item Consider impacts on privacy and informed consent
    \item Prioritize fairness over efficiency and convenience
    \item Balance transparency with security needs
\end{itemize}

 Remember: Technology should serve human values, not the other way around.

\rule{\linewidth}{0.4pt}

\textbf{GROUP B (Control) - Training Material:}
\rule{\linewidth}{0.4pt}
MODULE 4 (4/5)

DATA BACKUP: Protecting Your Information 

Regular backups safeguard your data against loss or ransomware. 

Effective backup strategies include: 
\begin{itemize}
    \item Maintain at least three copies of important data
    \item Store backups in different physical locations
    \item Test your backup recovery process periodically
\end{itemize}

Remember: The question isn't if data loss will occur, but when - be prepared.

\rule{\linewidth}{0.4pt}

\subsubsection{Module 5: Transparency}

\textbf{GROUP A (AIJET) - Training Material:}
\rule{\linewidth}{0.4pt}
MODULE 5 (5/5)
 
TRANSPARENCY: Document Your Process

Transparency means clearly documenting how and why decisions were made. 

For transparent decision-making:
- Record the specific steps you took to verify information
- Document which sources were consulted and when
- Share your verification methods with relevant stakeholders

Remember: Showing your work builds trust and enables learning from experience.

\rule{\linewidth}{0.4pt}

\textbf{GROUP B (Control) - Training Material:}
\rule{\linewidth}{0.4pt}
MODULE 5 (5/5)

NETWORK SECURITY: Protecting Your Connection 

Secure networks help prevent unauthorized access to your data. 

Network security best practices include:
\begin{itemize}
    \item Use strong encryption (WPA3) for all Wi-Fi networks
    \item Change default router passwords immediately
    \item Use a VPN when connecting to public Wi-Fi
\end{itemize}

Remember: Your network is only as secure as its weakest configuration setting.

\rule{\linewidth}{0.4pt}

\subsection{Questionnaire Items}

\begin{table*}[h]
\centering
\footnotesize
\caption{Awareness Assessment Results}
\begin{tabular}{p{6cm}ccc}
\toprule
\textbf{Statement} & \textbf{AIJET Group} & \textbf{Control Group} & \textbf{AIJET Effect} \\
\midrule
NASA confirms discovery of liquid water on Pluto in radar data released 5 May 2025. & 53.9\% & 54.7\% & -0.8\% \\
Scientists in Japan develop bamboo-based battery with 70\% longer lifespan than lithium-ion. & 38.2\% & 41.3\% & -3.1\% \\
A newly discovered penguin colony in Antarctica houses exactly 2,523 breeding pairs according to satellite imagery. & 53.9\% & 48.0\% & +5.9\% \\
The Hubble Space Telescope was launched into orbit in April 1990 aboard the Space Shuttle Discovery. & 68.4\% & 70.7\% & -2.3\% \\
Scientists at Oxford University discovered a deep-sea fish that can change its color precisely 37 times per minute as a defense mechanism. & 55.3\% & 50.7\% & +4.6\% \\
The UN General Assembly votes 121-14 on global plastics treaty draft. & 82.9\% & 70.7\% & +12.2\% \\
The final episode of the TV show `Friends' aired in 2004 and was watched by 52.5 million viewers in the US. & 68.4\% & 57.3\% & +11.1\% \\
AAAI, the American Association for Artificial Intelligence, has been retroactively credited with the discovery of America. & 75.0\% & 66.7\% & +8.3\% \\
\bottomrule
\end{tabular}
\label{tab:awareness_results}
\end{table*}

\begin{table*}[h]
\centering
\footnotesize
\caption{Integrity Assessment Results}
\begin{tabular}{p{6cm}ccc}
\toprule
\textbf{Statement with Source Image} & \textbf{AIJET Group} & \textbf{Control Group} & \textbf{AIJET Effect} \\
\midrule
The 2024 FBI IC3 report shows seniors lost \$4 billion to fraud. 

    \includegraphics[width=1\linewidth]{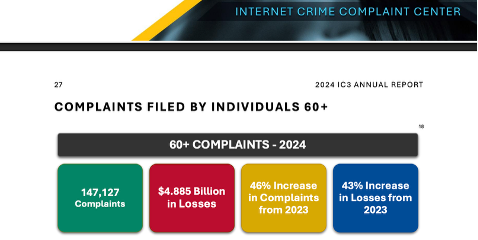}

& 30.3\% & 16.0\% & +14.3\% \\
The Women in AI Ethics™ organization in the US publishes a list of `50 Brilliant Women in AI Ethics' annually.

    \includegraphics[width=1\linewidth]{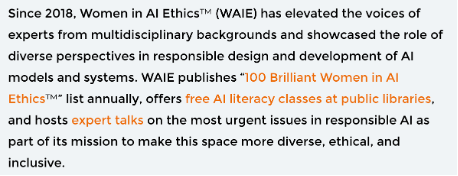}

& 51.3\% & 37.3\% & +14.0\% \\
The average American spends 7.5 hours daily on digital devices.

    \includegraphics[width=1\linewidth]{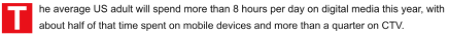}

& 40.8\% & 32.0\% & +8.8\% \\
Over 60\% of companies use AI in their cybersecurity operations.

    \includegraphics[width=1\linewidth]{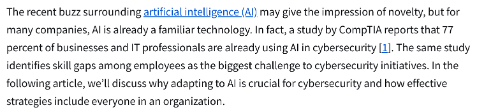}

& 80.3\% & 80.0\% & +0.3\% \\
Sam Altman, co-founder of ChatGPT, has won in 2025 the Turing Award.

    \includegraphics[width=1\linewidth]{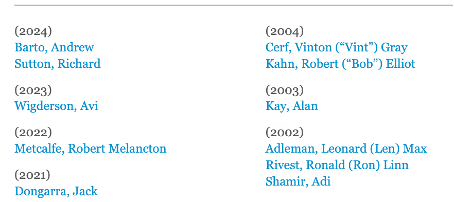}

& 69.7\% & 52.0\% & +17.7\% \\
\bottomrule
\end{tabular}
\label{tab:integrity_results}
\end{table*}

\begin{table*}[h]
\centering
\small
\caption{Judgment Assessment Results}
\begin{tabular}{p{4cm}p{4cm}ccc}
\toprule
\textbf{Email Scenario} & \textbf{Best Response} & \textbf{AIJET Group} & \textbf{Control Group} & \textbf{AIJET Effect} \\
\midrule
Finance department urgent vendor form request & Contact Finance through official channels to verify & 85.5\% & 84.0\% & +1.5\% \\
CEO requesting Amazon gift cards via Gmail & Check with CEO in person or via official channels & 80.3\% & 65.3\% & +15.0\% \\
IT support password reset link & Check with IT through official channels & 88.2\% & 78.7\% & +9.5\% \\
\bottomrule
\end{tabular}
\label{tab:judgment_results}
\end{table*}

`
\begin{table*}[h]
\centering
\footnotesize
\caption{Ethical Responsibility Assessment - Module 4}
\begin{tabular}{p{12cm}p{4cm}}
\toprule
\textbf{Assessment Component} & \textbf{Details} \\
\midrule
\textbf{Scenario} & Your company has started using an AI tool that analyzes email behavior. The tool has identified 8 employees as ``high-risk'' for clicking on phishing emails. The HR department wants to publish these employees' names on the company intranet as a security warning. \\
\midrule
\multirow{5}{*}{\textbf{Response Options}} & \textbf{A} - Seek consent from the employees before publishing any names \\
& \textbf{B} - Publish the list but make it anonymous (e.g., ``8 employees from various departments'') \\
& \textbf{C} - Proceed with publishing the full list to maximize security awareness \\
& \textbf{D} - Replace public naming with mandatory training for those identified \\
& \textbf{E} - Use the data to justify monitoring those employees' emails more closely \\
\midrule
\textbf{Instructions} & Select up to TWO actions you think are most ethical \\
\midrule
\textbf{Correct Responses} & A, B, or D selected \\
\midrule
\multirow{3}{*}{\textbf{Results}} & AIJET Group: 52.0\% \\
& Control Group: 36.0\% \\
& AIJET Effect: +16.0\% \\
\bottomrule
\end{tabular}
\label{tab:ethics_assessment}
\end{table*}

\begin{table*}[h]
\centering
\footnotesize
\caption{Transparency Assessment - Module 5}
\begin{tabular}{p{12cm}p{4cm}}
\toprule
\textbf{Assessment Component} & \textbf{Details} \\
\midrule
\textbf{Question} & Regarding the Amazon-gift-card email evaluated earlier, which step would you document first in your incident notes? \\
\midrule
\multirow{6}{*}{\textbf{Response Options}} & \textbf{A} - Record the second-channel verification (e.g., call or chat with the CEO via official contact) \\
& \textbf{B} - Note that gift-card requests are often scams, so I deleted the message \\
& \textbf{C} - Capture a screenshot of the email for the security team \\
& \textbf{D} - Flag the message as ``phishing'' in my mail client and archive it \\
& \textbf{E} - Forward the email to colleagues to ask if anyone else received it \\
& \textbf{F} - I would not record any notes for this request \\
\midrule
\textbf{Correct Responses} & A, C, or D selected \\
\midrule
\multirow{3}{*}{\textbf{Results}} & AIJET Group: 84.2\% \\
& Control Group: 81.3\% \\
& AIJET Effect: +2.9\% \\
\bottomrule
\end{tabular}
\label{tab:transparency_assessment}
\end{table*}

\end{document}